\documentstyle[prb,aps]{revtex}           

\begin{document}

\title{A derivation of the homogenous Maxwell equations}

\author{C\u alin Galeriu}

\address{Department of Physics, Clark University, Worcester, MA 01610, USA}

\date{\today}

\maketitle

\begin{abstract}
We present a theoretical derivation of the homogenous Maxwell equations, 
based on Stokes theorem for Minkowski space tensors.
\end{abstract}


\section{Introduction}
The Maxwell equations (1)-(2) for the electromagnetic field and the 
Lorentz 4-force law (3) for a charged particle are generalizations 
based on the experiments on the forces between electric charges and 
currents. These equations can be written in the covariant form 
\cite{jackson,griffiths}
\begin{equation}
{\partial {\cal F}^{\alpha\beta} \over x^{\alpha}} = {4 \pi \over c} 
J^{\beta},
\end{equation}
\begin{equation}
{\partial {\cal F}_{\alpha\beta} \over \partial x_{\gamma}} +
{\partial {\cal F}_{\beta\gamma} \over \partial x_{\alpha}} +
{\partial {\cal F}_{\gamma\alpha} \over \partial x_{\beta}} = 0,
\end{equation} 
\begin{equation}
F^{\alpha} = {q \over c} {\cal F}^{\alpha\beta} U_{\beta},
\end{equation}
\noindent where $x$ is the position 4-vector, $J$ is the 4-current, 
$F$ is the 4-force, $U$ is the 4-velocity, and ${\cal F}$ is the 
antisymmetric field-strength tensor.

These equations are intimately related with the principles of special 
relativity. Indeed, it was the consistent treatment of the electrodynamics 
of moving bodies that led to relativity \cite{einstein}. 
Tolman \cite{tolman} and Jefimenko \cite{jefimenko} have derived the 
Lorentz force law from the Maxwell equations and special relativity. 
Frisch and Wilets \cite{frisch} have derived the Maxwell equations and 
the Lorentz force law by applying the transformations of special 
relativity to Gauss's law of the flux of the electric field. 
Dyson \cite{dyson} reproduces an argument due to Feynman, in which
Maxwell equations follow from 
Newton's law of motion and the quantum mechanics commutation relations.
It is remarkable how nonrelativistic assumptions can lead to 
relativistically invariant equations. In this paper we derive 
the homogenous Maxwell equations by using Stokes theorem, in a fully
relativistic manner.

\section{Derivation of the homogenous Maxwell equations}
A consequence of the antisymmetry of the field-strength tensor is that 
the magnitude $i m_o c$ of the 4-momentum $p$ is constant. We will 
consider an extension of the Lorentz 4-force law to the case of a 
field-strength tensor not necessarily antisymmetric. Consequently the 
rest mass $m_o$ of the test particle will no longer be constant, but 
the electric charge will not be modified. We will require that the rest 
mass form a scalar field, since it is unphysically to assume that the 
rest mass of a particle at a given SpaceTime point might depend on the 
history of that particle. This is the underlying physical principle 
behind the homogenous Maxwell equations. The real-world situation is 
obviously a special case of this more general theory, limited to
an antisymmetric field-strength tensor. 

Consider two SpaceTime events A and B, and a charged test particle 
moving from A to B on any possible smooth path $\Gamma$, restricted 
only to the condition that the initial and final velocities be given. 
Since at A and B the direction of the 4-momentum is given, and the 
magnitude of the 4-momentum is also unique, we can conclude that the 
variation of the 4-momentum between A and B is the same regardless of 
the path followed. For two different paths, $\Gamma_1$ and $\Gamma_2$, 
we can write
\begin{equation}
p^{\alpha}_B - p^{\alpha}_A = \int_{\Gamma_1} dp^{\alpha} = 
\int_{\Gamma_2} dp^{\alpha}.
\end{equation}
The expression (3) of the Lorentz 4-force allows us to write the 
integrals in (4) as the circulation of the field-strength tensor
\begin{equation}
\int_{\Gamma} dp^{\alpha} = \int_{\Gamma} F^{\alpha} d\tau = 
\int_{\Gamma} {q \over c} {\cal F}^{\alpha\beta} {dx_{\beta} 
\over d\tau} d\tau = \int_{\Gamma} {q \over c} {\cal F}^{\alpha\beta} 
dx_{\beta}.
\end{equation}
By collecting the integrals in (4) on one side, and using (5), we obtain 
that in general
\begin{equation}
\oint {q \over c} {\cal F}^{\alpha\beta} dx_{\beta} = 0 
\Rightarrow \oint {\cal F}^{\alpha\beta} dx_{\beta} = 0.
\end{equation}

Stokes theorem, usually used in connection with the null circulation 
of a vector, will now be applied for the more general case of a tensor. 
Stokes theorem in the 4-dimensional Minkowski-space takes the form 
\cite{landau}
\begin{equation}
\oint {\cal F}_{\alpha\beta} dx_{\beta} = 
{1 \over 2} \int df_{\beta\gamma} 
({\partial {\cal F}_{\alpha\gamma} \over \partial x_{\beta}} - 
{\partial {\cal F}_{\alpha\beta} \over \partial x_{\gamma}}),
\end{equation}
where $df_{\beta\gamma}$ are projections of a surface element. Due to 
the arbitrary nature of the paths $\Gamma_1$ and $\Gamma_2$, from 
equations (6)-(7) it follows that
\begin{equation}
{\partial {\cal F}_{\alpha\gamma} \over \partial x_{\beta}} - 
{\partial {\cal F}_{\alpha\beta} \over \partial x_{\gamma}} = 0.
\end{equation}
This is the most general condition that the field-strength tensor must 
satisfy. We separate the symmetric and the antisymmetric components 
in (8), obtaining
\begin{equation}
{\partial {\cal F}_{\gamma\alpha}^{(s)} \over \partial x_{\beta}} - 
{\partial {\cal F}_{\alpha\beta}^{(s)} \over \partial x_{\gamma}} = 
{\partial {\cal F}_{\gamma\alpha}^{(a)} \over \partial x_{\beta}} + 
{\partial {\cal F}_{\alpha\beta}^{(a)} \over \partial x_{\gamma}}.
\end{equation}
From (9) we obtain two more equations by cyclic permutations of the 
indices ($\alpha \to \beta, \beta \to \gamma, \gamma \to \alpha$). By 
summing up all the three equations the symmetric components cancel, and 
we end up with the homogenous Maxwell equations
\begin{equation}
{\partial {\cal F}_{\alpha\beta}^{(a)} \over \partial x_{\gamma}} +
{\partial {\cal F}_{\beta\gamma}^{(a)} \over \partial x_{\alpha}} +
{\partial {\cal F}_{\gamma\alpha}^{(a)} \over \partial x_{\beta}} = 0.
\end{equation}

Due to the nonmetrical nature of Stokes theorem \cite{synge}, this 
derivation of the homogenous Maxwell equations will work in general 
relativity too.

\end{document}